# Counterintuitive Reconstruction of the Polar O-Terminated ZnO Surface With Zinc Vacancies and Hydrogen

**Ryan Jacobs**[1,†]**, Bing Zheng**[2,†]**, Brian Puchala**[3]**, Paul M. Voyles**[1]**, Andrew B. Yankovich**[4]**, and Dane Morgan**[1,*]

[1]*Department of Materials Science and Engineering, University of Wisconsin - Madison, Madison, Wisconsin 53706, USA*
[2]*School of Materials Science and Engineering, Beijing Institute of Technology, No. 5 Yard, Zhong Guan Cun South Street, Haidian District, Beijing 100081, People's Republic of China*
[3]*University of Michigan, 2300 Hayward St, Ann Arbor, Michigan 48109, USA*
[4]*Chalmers University of Technology, Gothenburg, Sweden SE-412 96*

[†]These authors contributed equally to this work

*Corresponding author: ddmorgan@wisc.edu

## Abstract

Understanding the structure of ZnO surface reconstructions and their resultant properties is crucial to the rational design of ZnO-containing devices ranging from optoelectronics to catalysts. Here, we are motivated by recent experimental work which showed a new surface reconstruction containing Zn vacancies ordered in a Zn(3×3) pattern in the subsurface of (0001)-O terminated ZnO. A reconstruction with Zn vacancies on (0001)-O is surprising and counterintuitive because Zn vacancies enhance the surface dipole rather than reduce it. In this work, we show using Density Functional Theory (DFT) that subsurface Zn vacancies can form on (0001)-O when coupled with adsorption of surface H and are in fact stable under a wide range of common conditions. We also show these vacancies have a significant ordering tendency and that Sb-doping created subsurface inversion domain boundaries (IDBs) enhances the driving force of Zn vacancy alignment into large domains of the Zn(3×3) reconstruction.

**Table of Contents Figure**

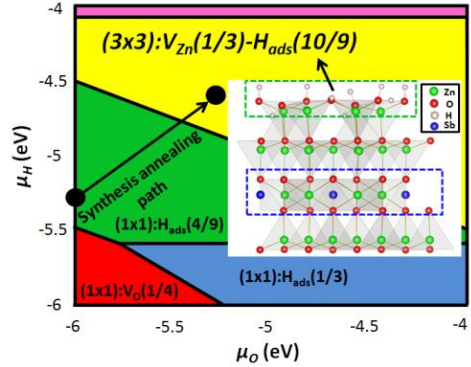

ZnO is a very promising material for applications in piezo-, micro-, and optoelectronic devices.[1] Recently, the substantial progress made in fabrication methods has led to the synthesis of a multitude of tailored ZnO surfaces and nanostructures with interesting and unique functional properties.[2-4]

The (0001)-oriented surfaces of the stable wurtzite structure of ZnO are Tasker type 3 polar surfaces which are electrostatically unstable, and this instability causes the surface to stabilize via a reconstruction.[5-6] For the (0001)-O terminated ZnO surface, surface charge neutrality arguments for common reconstructions require either the formation of O vacancies, adsorption of H, or larger scale formation of extended defects such as honeycomb-shaped pits.[7-10] In 2011, Wang *et al*[4] reported Sb-doped ZnO nanowires with p-type conductivity. Then, in 2012, Yankovich, *et al*[2] studied these nanowire surfaces in detail with Z-contrast scanning transmission electron microscopy (STEM). Although not discussed explicitly in Yankovich, *et al*[2], their studies showed a new surface reconstruction on the (0001)-O terminated surface with one column of Zn atoms missing along $\langle 10\bar{1}0 \rangle$ in the subsurface, as can be seen in the micrograph in **Figure 1**. The missing Zn atoms form a Zn vacancy ordered Zn(3×n) reconstruction. An ordered Zn(3x3) (i.e., n=3) reconstruction is the simplest surface reconstruction which explains the experimental observations such as the micrograph in Figure 1. While we cannot absolutely exclude larger ordered surfaces that are effectively supercells of the (3x3) cell, we have found that the (3x3) surface reconstruction is stable and constitutes the simplest reasonable physical model consistent with the experimental data. This work was the first time that Zn vacancies were found in the reconstruction of the (0001)-O surface. In addition to the Zn(3×3) reconstruction, Yankovich, *et al*[2] also found that Sb-decorated head-to-head basal plane inversion domain boundaries (IDBs) formed near the (0001)-O terminated growth surfaces of ZnO nanowires.[2] Above these IDBs the subsurface Zn atoms were arranged into an ordered (3×3) surface reconstruction commensurate with the Sb ordering in the IDB.

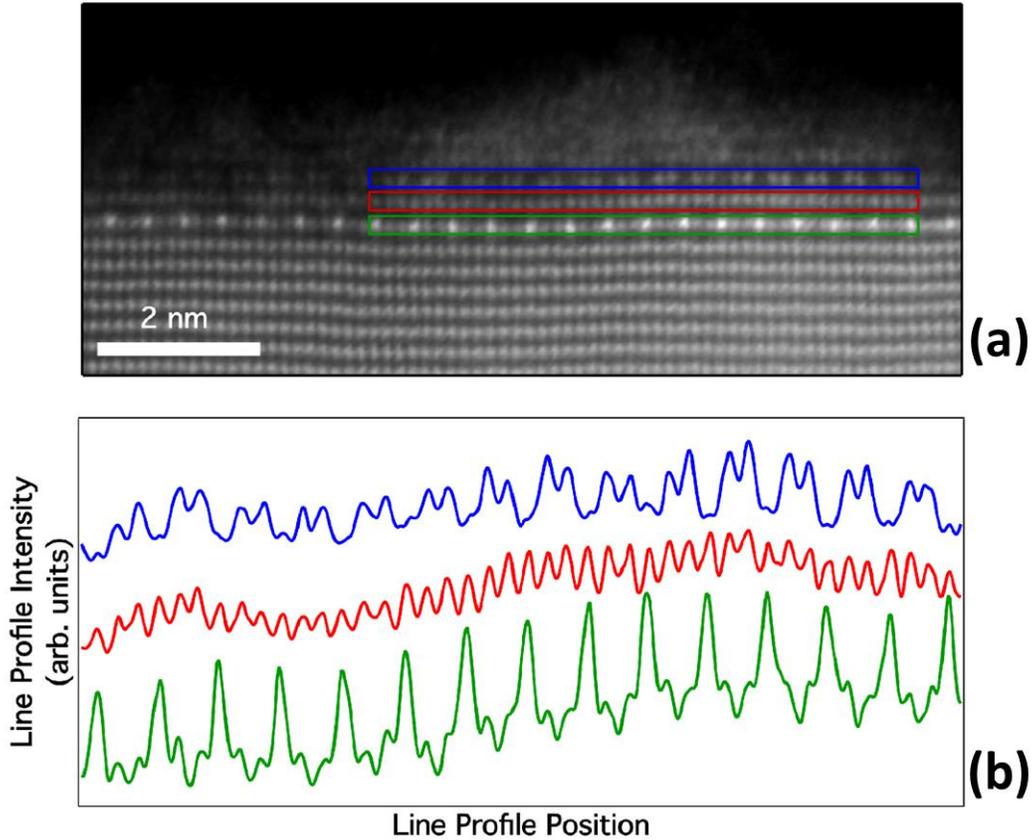

**Figure 1.** High angle annular dark field (HAADF) scanning transmission electron microscope (STEM) image of the Sb-doped ZnO nanowire growth surface along $\langle 10\bar{1}0 \rangle$. Due to the Z-contrast nature of HAADF STEM images, the Zn and Sb containing atomic columns have high contrast in the image, while the O and H are invisible. The Sb-decorated IDB, regular Zn, and Zn vacancy containing atomic planes are marked by the green, red, and blue boxes respectively. Intensity line profiles along each of the Sb-decorated IDB, regular Zn, and Zn vacancy containing atomic planes within the boxed region are shown in (b), revealing the correlated Sb and Zn vacancy atomic column structure. The image in (a) was originally published in Nano Letters in Ref. 2.

The experimental result of Yankovich, *et al*[2] suggested a new (0001)-O terminated surface reconstruction involving Zn vacancies was stable in the presence of the IDB, and that perhaps Zn vacancies play a role for surfaces without IDBs as well. However, the presence of Zn vacancies on the (0001)-O surface is surprising and counterintuitive because these defects enhance the surface polarity rather than reduce it. It is therefore important to determine if this reconstruction is actually an equilibrium structure of bulk

ZnO or if it is somehow specific to this material, perhaps a metastable surface formed during the nanowire synthesis or an otherwise unstable surface stabilized by the presence of the Sb-doped IDBs. A main result of the present work was that the ordered (3×3) subsurface Zn vacancy structure is in fact thermodynamically stable over a wide range of conditions, even without an Sb-doped IDB, provided that H adsorption is also considered. Our results further suggest that an intrinsic (3×3) Zn vacancy ordering tendency is enhanced by the Sb-doped IDB.

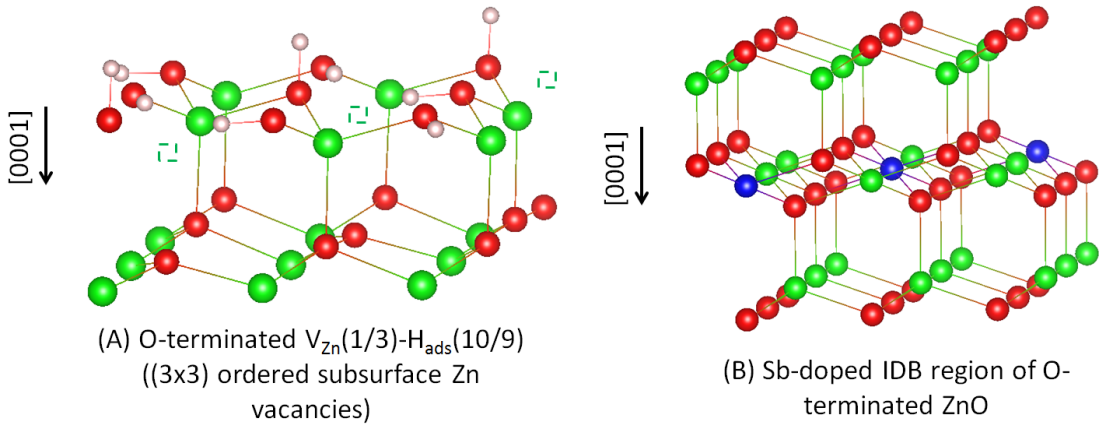

(A) O-terminated $V_{Zn}(1/3)$-$H_{ads}(10/9)$ ((3×3) ordered subsurface Zn vacancies)

(B) Sb-doped IDB region of O-terminated ZnO

**Figure 2.** ZnO stable surface reconstruction and geometry of Sb-doped IDB. (a) The O-terminated (3×3) Zn subsurface ordered vacancy structure with adsorption of 10 H per 9 surface O, (3×3): $V_{Zn}(1/3)$-$H_{ads}(10/9)$ and (b) the Sb-doped IDB region below the top Zn-O bilayer of an O-terminated surface. The green, red, white and blue spheres represent Zn, O, H and Sb atoms, respectively. The red and green dashed boxes represent O and Zn vacancies, respectively.

This work used Density Functional Theory (DFT) to explore the stability of Zn vacancies in the (0001)-O terminated ZnO surface and their ordering in the presence of a near-surface IDB. Specifically, we studied the stable surface reconstructions of the (0001)-O surface with and without the Sb-doped IDB in the presence of surface O vacancies, subsurface Zn vacancies, adsorbed H, and adsorbed OH as a function of the O and H chemical potentials, $\mu_O$ and $\mu_H$. The values of $\mu_O$ and $\mu_H$ are governed by the environmental variables temperature $T$, hydrogen partial pressure $P(H_2)$ and oxygen partial pressure $P(O_2)$.

All DFT calculations were performed using the VASP code[11-12] and GGA-PBE functionals[13] with the Hubbard $U$ correction[14-15] ($U$-$J$ = 7.5 eV).[16] Additional details on the DFT calculation methods and slab calculations can be found in **Section A** and **Section B** of the **SI**, respectively. The Zn vacancy ordered Zn(3×3) reconstruction (with adsorbed H) investigated in this work ((3×3): $V_{Zn}(1/3)$-$H_{ads}(10/9)$) is shown in **Figure 2(a).** The structure of the IDB region beneath the top surface Zn-O bilayer is shown in **Figure 2(b)**. Additional details on the structural characteristics of the Sb-doped IDB are provided **Section D** of the **Supplementary Information (SI)**. We have explored an extensive set of potential ZnO (0001)-O surface reconstructions for determining phase stability. Generally, the thermodynamically stable ZnO surface reconstructions are those that contain defects and/or adsorbed atoms to sufficiently compensate the surface polarity,[5-6, 17] thus producing charge neutral surfaces.[7, 9, 18] Although we have not performed a comprehensive search of all possible surface reconstructions, to the best of our knowledge all ZnO (0001)-O surface reconstructions in the literature that have been observed experimentally or predicted as stable by theory have been considered, as well as a number of new structures inspired by the (3×3) structures observed by Yankovich, *et al.*[2] In addition to the (0001)-O surface, there have been numerous studies on the complimentary (0001)-Zn surface, a few of which we mention here. For the (0001)-Zn surface, both scanning tunneling microscopy and first-principles calculation have demonstrated that terraces with a triangular shape and a step height of one ZnO double layer can occur.[18-19] Torbrügge *et al.*[20] also reported a Zn(1×3) reconstruction on ZnO (0001)-Zn surface based on scanning force microscopy. Lastly, Li *et al.* used ab initio molecular dynamics to explore a new (0001)-Zn reconstruction resulting from an O extrusion process.[21]

To describe our set of ZnO (0001)-O surface structures more specifically, we note that two distinct sets of surfaces were investigated. The first set consisted of pure (2×2) and (3×3) (0001)-O ZnO surfaces. The second set consisted of (3×3) (0001)-O ZnO surfaces with an Sb-doped IDB as shown in **Figure 2(b)**. The (2×2) and (3×3) reconstructions were investigated with slab geometries using 2×2 and 3×3 supercells of the ZnO conventional cell. However, in order to investigate the (2×2) reconstructions and the Sb-IDB (which forms a (3×3) ordering) in the same cell we used a larger (6×6) cell. For all surface calculations, O vacancies, H adsorption, and combined subsurface Zn vacancies with H adsorption were used to produce surfaces which were either charge neutral or deviated slightly from charge neutrality because of stoichiometry constraints from the size of the finite unit cells. Additionally, combined OH adsorption with Zn vacancies was also considered, as well as the hexagonal pit reconstructions investigated by Wahl *et al.*[9] It is important to note that we did not calculate the hexagonal pit reconstructions with the IDB present. These hexagonal reconstructions with the IDB were not simulated because the formation of the hexagonal pits requires movement of the subsurface Zn atoms which comprise the Sb-doped IDB, thus making the simulation cells both very complex and significantly different from the geometry of the hexagonal pits in ZnO without the IDB. However, these phases are not expected to be stabilized by the IDB and therefore no attempt was made to establish the lowest energy structure of the hexagonal-type reconstruction in the presence of the IDB. All surface structures considered are summarized in **Section B** of the **SI.** In all cases, the complementary non-reconstructed surface on the other side of the slab was passivated with a monolayer of pseudo-H atoms such that the surface polarity was compensated.[22-23]

The DFT surface reconstruction energies were combined with thermodynamic phase stability analysis to determine the (0001)-O ZnO surface phase diagrams with and without the IDB. More specifically, the phase diagrams were constructed by calculating the surface energy of all surface reconstructions as a function of the O and H chemical potentials, $\mu_O$ and $\mu_H$, and the stable surface reconstruction is the one with the lowest surface energy for a given $\mu_O$ and $\mu_H$. In addition to O and H, $H_2O$ can be an important reactant to consider for ZnO surface reconstructions, such as the $(10\bar{1}0)$ surface.[24] However, in our present study, we do not explicitly include surface reconstructions of (0001)-O ZnO with $H_2O$ because it has been shown experimentally that molecular $H_2O$ does not adsorb on the (0001)-O surface under the conditions relevant in this study. For instance, temperature desorption spectroscopy measurements have shown that no $H_2O$ is present on the (0001) O-terminated surface above approximately room temperature conditions, even under high fluxes of $H_2O$.[25] In addition, the work of Kunat, *et al.*[26] has shown using He-ion scattering, low energy electron diffraction and X-ray photoelectron spectroscopy that $H_2O$ does not adsorb on the (0001) O-terminated ZnO surface as molecular $H_2O$, but instead decomposes and forms O-H bonds with the surface. These decomposed $H_2O$ molecules which form O-H bonds create a partially H-terminated surface which, for a (2x2) ZnO surface cell, is indistinguishable from the H(1x1) reconstruction containing roughly 50% H adsorption on the (0001) O-terminated surface. More information on the thermodynamic methods used here can be found in **Section C** of the **SI**.

**Figure 3** depicts the surface reconstruction phase diagram as a function of $\mu_O$ and $\mu_H$, and summarizes the essential results of our study on the (0001)-O ZnO surface. The secondary axes of **Figure 3** tabulate the values of $P(O_2)$ and $P(H_2)$ for $T = 500, 700, 900$ and 1150 K. It should be noted that in this work only select compositions have been investigated, and that the exact stoichiometries in experiment may therefore differ from those studied. Therefore, the stable phases identified in the calculations should be taken as approximations to the true composition. Higher values of $\mu_O$ ($\mu_H$) correspond to more oxidizing (reducing) conditions. Thus, the lower right corner of **Figure 3** corresponds to the most oxidized surfaces and the upper left corresponds to the most reduced surfaces.

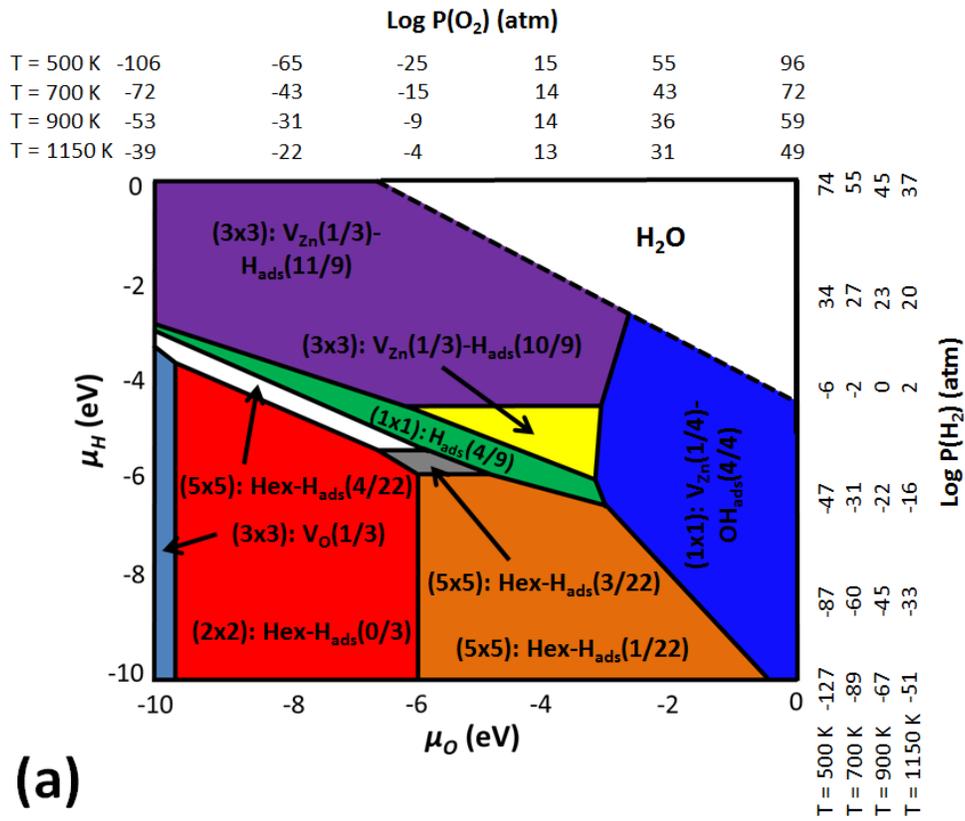

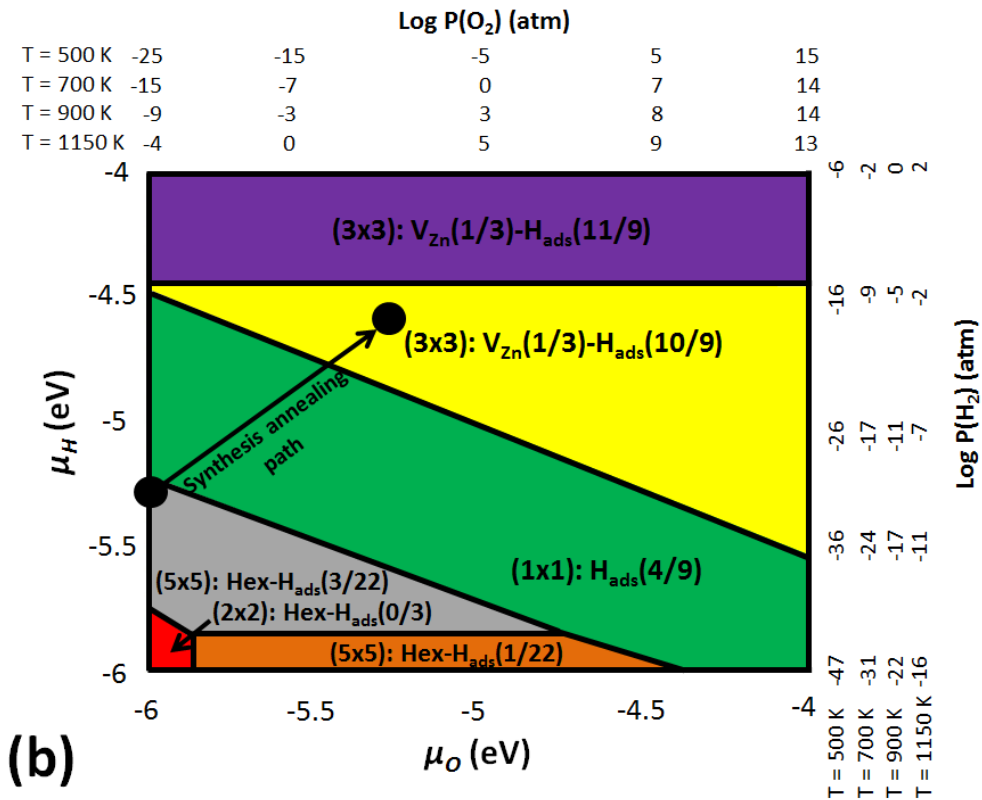

**Figure 3.** Phase diagrams summarizing the stable surfaces of (0001)-O ZnO. The phase diagram in (a) depicts the stable surfaces for a wide range of $\mu_O$ and $\mu_H$. The phase diagram in (b) depicts the stable surfaces expected over the $\mu_O$ and $\mu_H$ values relevant for most experimental syntheses of ZnO. In both (a) and (b), the secondary axes provide values of $P(O_2)$ and $P(H_2)$ for different temperatures which equate to the specific value of $\mu_O$ and $\mu_H$. In (a) the region of stability of $H_2O$ in equilibrium with $O_2$ and $H_2$ is filled in white. The nomenclature employed here to denote the surface phase composition is in the form of: $V_N(n/x)$-$M_{ads}(m/y)$, where $V_N$ denotes vacancies of species $N$ ($N$ = O or Zn), with a concentration of $n$ vacancies per $x$ surface $N$ atoms and $M_{ads}$ denotes $M$ adsorption with a concentration of $m$ $M$ atoms per $y$ surface O atoms. The nomenclature of $Hex$-$H_{ads}(m/y)$ indicates the hexagonal pit reconstruction following Lauritsen *et al*[8] and Wahl *et al*[9], where the $m$ value denotes the number of adsorbed hydrogen per $y$ surface O atoms. The annealing pathway during ZnO nanowire synthesis is indicated in (b) with the black arrow between the beginning and ending conditions showing that synthesized nanowires from Yankovich *et al*[2] will exhibit the ordered (3×3):$V_{Zn}(1/3)$-$H_{ads}(10/9)$ reconstruction. The positions of the synthesis conditions on the phase diagram are further discussed in the main text and **Section E** of the **SI**.

The results of our phase diagram in **Figure 3** are in good agreement with known experimental results and previous predictions. Specifically, below 600 K, the (0001)-O surface has a significant affinity to H atoms, causing the O vacancies to fill and adsorption of ≈44% H ($H_{ads}(4/9)$), resulting in the (1×1): $H_{ads}(4/9)$ reconstruction for temperatures below 600 K for a wide range of synthesis relevant $\mu_O$ and $\mu_H$.[27] In recent years, a number of further studies have been performed on the H configuration[7, 10] and O reconstruction[8] on the (0001)-O terminated surface of ZnO. H coverage on the O-terminated surface of ZnO was demonstrated to be in the range 33%~50% from low-energy electron diffraction (LEED)[10] and Density Functional Theory (DFT) calculations[7, 10], values in good agreement with our predicted H coverages on the (1×1) reconstruction. In addition, scanning tunneling microscopy (STM) and DFT studies by Lauritsen, *et al*[8] and Wahl, *et al*[9] revealed the presence of (2×2) and (5×5) hexagonal pit reconstructions under ultra-high vacuum conditions. In this work, we find that these hexagonal pit reconstructions are stable under low $P(H_2)$ conditions, in agreement with Wahl*, et al*. However, we find that at higher $P(H_2)$ the hexagonal reconstructions become unstable and instead the (3×3):$V_{Zn}(1/3)$-$H_{ads}(10/9)$ ordered subsurface Zn vacancy structure observed by Yankovich, *et al*[2] is the stable surface structure.

**Figure 3(b)** shows the stable structures present under typical ZnO synthesis and annealing conditions. In particular, the two black circles and black line illustrate the initial synthesis condition, final annealed condition, and approximate path of annealing that were used by Yankovich, *et al*[2] (the chemical potentials for this path in **Figure 3(b)** and given in the following discussion are derived in **Section E** of the **SI**).[2, 4, 28] At T = 1150 K, $\mu_O$ = -6.00 eV/O and $\mu_H$ = -5.37 eV/H, the (5×5): Hex-$H_{ads}(3/22)$ structure is stable. However, during annealing the system cools and moves to higher $\mu_O$ and $\mu_H$. At around 700 K, $\mu_O$ = -5.26 eV/O and $\mu_H$ = -4.54 eV/H the system stabilizes (3×3): $V_{Zn}(1/3)$-$H_{ads}(10/9)$, which is the ordered phase observed by Yankovich, *et al.*[2] As the cations are likely still mobile at 700 K this result suggests that Zn vacancies have a driving force to order in the Zn(3×3) reconstruction even without the Sb-doped IDB present.

The corresponding phase diagram to that of **Figure 3** showing stable surface reconstructions with the Sb-doped IDB present, is given in **Figure SI-2** in **Section F** of the **SI**. When the Sb-doped IDB is present and under the final annealed condition of 700 K, $\mu_O$ = -5.26 eV/O and $\mu_H$ = -4.54 eV/H, the same equilibrium surface reconstruction of (3×3): $V_{Zn}(1/3)$-$H_{ads}(10/9)$ seen in **Figure 3** for pure (no IDB) ZnO is stable. However, when under conditions of the onset of annealing of 1150 K, $\mu_O$ = -6.00 eV/O and $\mu_H$ = -5.37 eV/H, the (1×1): $H_{ads}(4/9)$ reconstruction is stable, which is different from the stable (5×5): Hex-$H_{ads}(3/22)$ structure seen under these conditions in **Figure 3**. However, as noted above, we are not able to model the hexagonal phases in the presence of the IDB so we cannot say whether one should expect some structure analogous to the hexagonal pit reconstructions to form under conditions of very low $\mu_O$ and $\mu_H$ when the Sb-doped IDB is present. The possible presence of the hexagonal phases with the IDB under some conditions does not impact the main conclusions of this work, which is that the (3×3): $V_{Zn}(1/3)$-$H_{ads}(10/9)$ surface is stable without the IDB.

Based on our results in **Figure 3** and **Figure SI-2**, (0001)-O ZnO is predicted to form ordered Zn vacancies with adsorbed H under the experimental annealing conditions during cooling, regardless if an Sb-doped IDB is present or absent. However, it is reasonable to expect that the IDB might couple to the vacancies, and we did find that the IDB helps order the vacancies, although it is not critical to stabilizing them. We demonstrated how the Sb-doped IDB helps the Zn vacancy ordering by comparing the energies between structures that have $V_{Zn}(1/3)$ in the subsurface Zn layer but different vacancy orderings, one with (3×3) ordering aligned above the Sb atoms and three with vacancies randomly distributed. The energy to form the ordered Zn vacancy arrangement above the IDB was 0.25 eV/(Zn vacancy), while the average disordered formation energy was 0.80 eV ± 0.13 eV/(Zn vacancy) (where 0.13 eV/(Zn vacancy) is the standard error in the mean). The structure with ordered vacancies above the Sb sites is therefore 0.55 ± 0.13 eV/(Zn vacancy) more stable than the average energy of structures with random vacancies. For the (0001)-O surface without an IDB the same energy difference is 0.37 ± 0.05 eV/(Zn vacancy). This result suggests that Zn vacancies will have a driving force to order in the (3×3) pattern whether or not an Sb-doped IDB is present, but that the presence of the Sb-doped IDB significantly strengthens the driving force for ordering. Furthermore, if the ordered Zn vacancies are shifted such that they are not vertically aligned with the underlying Sb atoms of the IDB, the formation energy is 0.47 eV/(Zn vacancy) higher than when the vacancies are ordered directly above the Sb-doped IDB. This result indicates that while there is a generally strong driving force for Zn vacancies to order along the basal plane diagonal (with or without Sb-doped IDB), there is an additional driving force for the Zn vacancies to order directly above the Sb-doped IDB when an IDB is present. To the extent that the IDB domains are large, this alignment will suppress the formation of many small, poorly aligned ordered domains, with lines of Zn vacancies pointing in different directions, and support the development of large regions of ordered domains aligned with the IDB. This alignment might enhance the Zn vacancy ordering compared to what would be observed without the IDB. We propose that the IDB drives alignment through strain and/or electrostatics, as the Sb are positively charged cations with a significantly different size than Zn, and could produce significant strain and/or electrostatic fields that propagate the ordering in the IDB plane up to the top layer of Zn vacancies.

In summary, this work demonstrates that a reconstruction for the (0001)-O surface which features Zn vacancies is thermodynamically stable provided a large amount of adsorbed H is also present to yield a zero surface charge (non-polar) surface reconstruction. In particular, this work has demonstrated that for (0001)-O terminated ZnO the (3×3): $V_{Zn}(1/3)$-$H_{ads}(10/9)$ surface is stable under a wide-range of relevant synthesis conditions. Furthermore, our results suggest that the Zn vacancies have a strong ordering tendency, in particular into the Zn(3×3) structure, and will tend to form ordered arrangements. We have shown that an Sb-doped IDB just below the surface is predicted to enhance the driving force to order the Zn vacancies in the Zn(3×3) pattern as well as align the vacancy pattern above the Sb atoms. In addition to predicting a new surface reconstruction of ordered Zn vacancies and adsorbed H, this work suggests that subsurface Zn vacancies are a general phenomenon in the growth, processing, and resultant electronic properties of oxygen terminated ZnO which deserve further attention.

**ACKNOWLEDGMENTS**

Electron microscopy characterization and initial exploration of stable Zn surface structures was supported by the Department of Energy, Basic Energy Sciences (DE-FG02-08ER46547). Electron microscopy experiments used facilities and instrumentation supported by NSF through the University of Wisconsin Materials Research Science and Engineering Center (DMR-1121288). Calculations benefitted from the use of parallel supercomputing resources provided by the Extreme Science and Engineering Discovery Environment (XSEDE), supported by National Science Foundation grant number OCI-1053575.

**Supporting Information Available**: Supporting information contains information on the DFT calculations, surface models, thermodynamic formalism, and phase diagram of ZnO containing the Sb-doped IDB. This material is available free of charge via the Internet at http://pubs.acs.org.